\documentclass[
    ,final            % use final for the camera ready runs
  ]
  {aipproc}

\layoutstyle{6x9}

\begin{document}

\title{Quaternionic and Octonionic Spinors}

\author{Francesco Toppan}{
  address={ CCP/CBPF, Rua Dr. Xavier Sigaud
150, cep 22290-180 Rio de Janeiro (RJ), Brazil}
}

\begin{abstract}
Quaternionic and octonionic spinors are introduced and their fundamental
properties (such as the space-times supporting them) are reviewed. The conditions 
for the existence of their associated Dirac equations are analyzed. 
Quaternionic and octonionic supersymmetric algebras defined in  terms of such spinors
are constructed. Specializing to the $D=11$-dimensional case, the relation of
both the quaternionic and the octonionic supersymmetries with the ordinary $M$-algebra are discussed.
\end{abstract}

\maketitle

%%%%%%%%%%%%%%%%%%%%%%%%%%%%%%%%%%%%%%%%%%%%
%% MAINMATTER
%%%%%%%%%%%%%%%%%%%%%%%%%%%%%%%%%%%%%%%%%%%%

\section{Introduction}

 The division algebras are responsible for many important mathematical structures
 of interest for physicists (as an example one can cite the Hopf fibrations).\par
 In this talk we review at first the well-known, see e.g. \cite{Por},
 \cite{Oku} connection between
 division algebras and Clifford algebras, explaining also in which sense we can 
 extend the (associative) notion of Clifford algebra in order to accommodate an alternative
 (i.e. non-associative) structure as the one given by the octonions. This motivates us to introduce quaternionic and octonionic spinors following \cite{CRT} for later studying their free dynamics (namely, their associated Dirac-type of equations) for any space-time supporting quaternionic or octonionic spinors. \par
 The potentially most interesting physical applications
 of the formalism here discussed concern supersymmetry. 
 There are several reasons for that. Division algebras, including quaternions and octonions, naturally enter the classification of supersymmetry, \cite{KT}. Octonions (for a review on them one can consult \cite{Bae}), which on mathematical side are the 
 most interesting structure (taking into account that they are the maximal division algebra
 and are associated with the existence of the exceptional Lie algebras, \cite{BS}),
 so far have not found any concrete application in physics, despite many attempts
 to introduce them in several different contexts (e.g. in order to explain the strong interactions
 and the confinement of quarks, \cite{GG}). On the other hand, see \cite{Boy} and references therein, one very appealing possibility could be found at the very heart of the unification program
 of the interactions which goes under the name of $M$-Theory. Indeed in the past \cite{FM}
 octonionic-valued superstrings have been described. More recently, \cite{LT1} and \cite{LT2}, it was shown
 that an octonionic-valued version of the $11$-dimensional $M$-algebra can be constructed
 and admits very peculiar properties. This opens the way to a possible octonionic formulation
 of the $M$-theory, which could correspond to some suggestions put forward in \cite{Boy} and \cite{Ram}.\par
 For what concerns the quaternionic supersymmetry, its mathematical aspects have been
 clarified and classified
 (see \cite{Top}). It must be said that no concrete
 physical implementation has yet been investigated. The most closely related relevant application so far concerns the analytic continuation of the $M$-algebra to the Euclidean \cite{LT3}. It is based on complex spinors living on a quaternionic spacetime.  
    
\section{On Clifford algebras and division algebras}

The basic relation defining a Clifford algebra is given by
\begin{eqnarray}\label{cliff}
\Gamma^\mu\Gamma^\nu+\Gamma^\mu\Gamma^\nu &=& 2\eta^{\mu\nu},
\end{eqnarray}
with $\eta^{\mu\nu}$ being a diagonal matrix of $(p,q)$ signature
(i.e. $p$ positive, $+1$, and $q$ negative, $-1$, diagonal
entries).
\par
On the other hand
the four division algebra of real (${\bf R}$) and complex (${\bf C}$) numbers, quaternions (${\bf H}$)
and octonions 
(${\bf O}$) possess respectively $0$, $1$, $3$ and $7$ imaginary elements $e_i$ satisfying the relations
\begin{eqnarray}
e_i\cdot e_j &=& -\delta_{ij} + C_{ijk} e_{k},
\label{octonrel}
\end{eqnarray}
($i,j,k$ are restricted to take the value $1$ in the complex case, $1,2,3$ in the quaternionic case and
$1,2,\ldots , 7$ in the octonionic case; furthermore, the sum over repeated indices is understood), with
$C_{ijk}$ the totally antisymmetric division-algebra structure constants. The octonionic division
algebra is the maximal, since quaternions, complex and real numbers can be obtained as its restriction. Its
totally antisymmetric octonionic structure constants can be expressed as
\begin{eqnarray}
&C_{123}=C_{147}=C_{165}=C_{246}=C_{257}=C_{354}=C_{367}=1&
\end{eqnarray}
(and vanishing otherwise).
\par
The octonions are the only non-associative, however alternative (see \cite{Bae}), division algebra.\par
It is therefore clear, due to the antisymmetry of $C_{ijk}$, that (\ref{cliff})
can be realized, for the
$(0,3)$ and the $(0,7)$ signatures, in terms of, respectively, the imaginary quaternions and the imaginary octonions.\par
With an abuse of language (due to their non-associativity) when in the following
we will speak about ``octonionic Clifford algebra" we will always have in mind the above connection.\par
For our later purposes it is of particular importance the notion of division-algebra principal conjugation.
Any element $X$ in the given division algebra can be expressed through the sum
\begin{eqnarray}
X&=&x_0 + x_ie_i ,
\end{eqnarray}
where $x_0$ and $x_i$ are real, the summation over repeated indices is understood and the positive integral
$i$ are restricted up to $1$, $3$ and $7$ in the ${\bf C}$, ${\bf H}$ and ${\bf O}$ cases respectively.
The principal conjugate $X^\ast$ of $X$ is defined to be
\begin{eqnarray}\label{conjug}
X^\ast&=&x_0 - x_ie_i .
\end{eqnarray}
It allows introducing the division-algebra norm through the product $X^\ast X$. The 
normed-one restrictions 
\begin{eqnarray}\label{normedone}
X^\ast X &=&1
\end{eqnarray}
 select the three parallelizable spheres $S^1$, $S^3$ and $S^7$
in association with ${\bf C}$, ${\bf H}$ and ${\bf O}$ respectively.
\par
Further comments on the division algebras and their relations with Clifford algebras can be found in \cite{CRT}.\par
The connection between division algebras and Clifford algebras can be extended to other
signature spacetimes as well.
The two following algorithms can be used to lift
$d$-dimensional Gamma matrices (denoted as $\gamma_i$) of
a $D=p+q$ spacetime with $(p,q)$ signature into
$2d$-dimensional $D+2$ Gamma matrices (denoted as $\Gamma_j$) of a
$D+2$ spacetime, produced according to either
\begin{eqnarray}
 \Gamma_j &\equiv& \left(
\begin{array}{cc}
  0& \gamma_i \\
  \gamma_i & 0
\end{array}\right), \quad \left( \begin{array}{cc}
  0 & {\bf 1}_d \\
  -{\bf 1}_d & 0
\end{array}\right),\quad \left( \begin{array}{cc}
  {\bf 1}_d & 0\\
  0 & -{\bf 1}_d
\end{array}\right)\nonumber\\
&&\nonumber\\ (p,q)&\mapsto&
 (p+1,q+1).\label{one}
\end{eqnarray}
or
\begin{eqnarray}
 \Gamma_j &\equiv& \left(
\begin{array}{cc}
  0& \gamma_i \\
  -\gamma_i & 0
\end{array}\right), \quad \left( \begin{array}{cc}
  0 & {\bf 1}_d \\
  {\bf 1}_d & 0
\end{array}\right),\quad \left( \begin{array}{cc}
  {\bf 1}_d & 0\\
  0 & -{\bf 1}_d
\end{array}\right)\nonumber\\
&&\nonumber\\ (p,q)&\mapsto&
 (q+2,p).\label{two}
\end{eqnarray}
The relation (\ref{cliff}) can therefore be realized, for specific spacetimes,
in terms of quadratic matrices with either quaternionic or octonionic entries.
The spacetimes supporting such Clifford realizations can be easily 
computed. In the octonionic case, up to $D=13$, we obtain the following 
list of octonionic spacetimes
\begin{eqnarray}\begin{array}{|c|c|}\hline
D=7 & (0,7), ~(7,0)\\ \hline D=8& (0,8), (8,0)\\ \hline D=9 &
(0,9), (9,0), (1,8), (8,1) \\ \hline D=10& (1,9), (9,1) \\ \hline
D=11 &(1,10), (10,1),(2,9),(9,2)
\\ \hline D=12& (2,10), (10,2) \\ \hline D=13& (3,10),(10,3),(2,11), (11,2)\\ \hline

\end{array}\label{octonspaces}
\end{eqnarray}
An analogous list can be produced in the quaternionic case too.
\subsection{A comment on the octonionic realization}

One should be aware of the properties of the non-associative
realizations of the relation (\ref{cliff}), in terms of Gamma-matrices
with octonionic-valued entries. In the octonionic case the
commutators 
\begin{eqnarray}\label{commuta}
\Sigma_{\mu\nu} &=&\relax [\Gamma_\mu, \Gamma_\nu]
\end{eqnarray}
no longer correspond, as in the associative case, to
the generators of the Lorentz group. They correspond
instead to the generators of the coset $SO(p,q)/G_2$, being $G_2$
the $14$-dimensional exceptional Lie algebra of automorphisms of
the octonions. This point can be easily illustrated with the
basic example of the Euclidean $7$-dimensional
case expressed by the imaginary octonions.  Their commutators give rise to $7=21-14$ generators,
isomorphic to the imaginary octonions. Indeed
\begin{eqnarray}
\relax [e_i,e_j]& = &2C_{ijk}e_k .\label{octcomm}
\end{eqnarray}
The alternativity property satisfied by the octonions implies that
the seven-dimensional commutator algebra among imaginary octonions
is not a Lie algebra, the Jacobi identity being replaced by a
weaker condition that endorses (\ref{octcomm}) with the structure
of a Malcev algebra (see \cite{Bae}).\par Such an algebra admits a
nice geometrical interpretation \cite{{LM},{CRT}}. Indeed, the
normed $1$ unitary octonions $X=x_0 + x_i e_i$
satisfying the (\ref{normedone}) condition 
describe the seven-sphere $S^7$. The latter is a parallelizable
manifold with a quasi (due to the lack of associativity) group
structure. \par
On the seven sphere,
infinitesimal homogeneous transformations which play the role of
the Lorentz algebra can be introduced through
\begin{eqnarray}
\delta X &=& a\cdot X,\end{eqnarray} with $a$ an infinitesimal
constant octonion. The requirement of preserving the unitary norm
(\ref{normedone}) implies the vanishing of the $a_0$ component, so
that $a \equiv a_ie_i$. Therefore, the above commutator algebra
(\ref{octcomm}), generated by the seven $e_i$, can be
interpreted as the algebra of ``quasi" Lorentz transformations
acting on the seven sphere $S^7$. At least in this specific
example we discovered a nice geometrical setting underlining the
use of the octonionic realization of the Clifford
relation (\ref{cliff}) with $(0,7)$ signature. Indeed, while the associative 
representation (realized by $8\times 8$ real matrices) of the
seven dimensional Clifford algebra is required for describing the
Euclidean $7$-dimensional flat space, the non-associative
realization describes the geometry of $S^7$.

\subsection{The Weyl condition}

Spinors can simply be introduced as column-vectors with entries valued in the
given division algebra and carrying a representation of the Lorentz-algebra
generators $\Sigma_{\mu\nu}$ introduced in (\ref{commuta}) (Octonionic
spinors, as discussed in the previous subsection, carry a representation of the 
$G_2$ coset).\par
A particular case arises for those space-time whose associated Gamma-matrices
can be
chosen to be block-antidiagonal, i.e.
of the form
\begin{eqnarray}
\Gamma^\mu &=&\left( \begin{array}{cc}
  0 & \sigma^\mu \\
  {\tilde\sigma}^\mu & 0
\end{array}\right)\label{weyl}
\end{eqnarray}
The corresponding signatures can be easily recovered from the introduced algorithmic
constructions (\ref{one}) and (\ref{two}). We will call the Gamma-matrices satisfying (\ref{weyl})
as ``generalized Weyl matrices". The generators $\Sigma_{\mu\nu}$ of (\ref{commuta})
in this case carry fundamental spinor-representations, realized by either upper or lower
Weyl spinors, whose number of components is only half of their size.\par
In the Weyl case two projectors $P_\pm$ can
be introduced through
\begin{eqnarray}
P_\pm &=&\frac{1}{2}({\bf 1}_{2d} \pm {\overline
\Gamma}),\nonumber\\ {\overline\Gamma} &=&\left( \begin{array}{cc}
  {\bf 1}_d & 0 \\
  0 & -{\bf 1}_d
\end{array}\right)\label{project}
\end{eqnarray}
and the corresponding chiral (upper components) and
antichiral (lower components) Weyl spinors,
whose number of components is half of the
size of the corresponding $\Gamma$-matrices, can be defined as satisfying
\begin{eqnarray}
\Psi_\pm &=& P_\pm \Psi.
\end{eqnarray} 

\section{Quaternionic and octonionic Dirac equations}

In the previous section we have introduced all the necessary ingredients
(division-algebra valued Clifford algebras and the associated spinors)
to define the quaternionic and octonionic versions of the Dirac equation. \par
In this section we introduce such equations and provide the full classification
\cite{CRT} of the spacetimes supporting them (and under which condition, i.e.
the possible presence of massive or pseudomassive terms, Weyl spinors, etc.).
The results will be presented in a series of tables. 
\par
Let us introduce at first the needed conventions. A matrix $A$, given by the product
of all time-like Gamma matrices and generalizing the role of $\Gamma^0$ in the Minkowskian 
case is used to define barred spinors (${\overline \psi}= \psi^\dagger
A$). (Pseudo)-kinetic and (pseudo)-massive terms can be introduced for
full ($NW$) and Weyl ($W$) spinors according to the following prescriptions
(in order to avoid unnecessary repetitions, it is sufficient to list here
the octonionic case,
the quaternionic one being easily recovered from the given formulas).\par  
The (pseudo)-kinetic terms are given according to
\begin{eqnarray}
 K_{X} &=&
\frac{1}{2}tr[(\Psi^{\dagger}_{} A \Gamma^{\mu}
X)\partial_{\mu} \Psi_{}]+
\frac{1}{2}tr[\Psi^{\dagger}_{} (A \Gamma^{\mu}
X\partial_{\mu} \Psi_{})], \nonumber \\ 
 K_{//X} &=&
\frac{1}{2}tr[(\Psi^{\dagger}_{+} A \Gamma^{\mu}
X)\partial_{\mu} \Psi_{+}]+
\frac{1}{2}tr[\Psi^{\dagger}_{+} (A \Gamma^{\mu}
X\partial_{\mu} \Psi_{+})], \nonumber \\ K_{\bot X}
&=&\frac{1}{2} tr[(\Psi^{\dagger}_{+} A
\Gamma^{\mu}X)\partial_{\mu} \Psi_{-}] +
\frac{1}{2}tr[\Psi^{\dagger}_{+} (A
\Gamma^{\mu}X\partial_{\mu} \Psi_{-})] +\nonumber\\ &&
\frac{1}{2} tr[(\Psi^{\dagger}_{-} A
\Gamma^{\mu}X)\partial_{\mu} \Psi_{+}] +
\frac{1}{2}tr[\Psi^{\dagger}_{-} (A
\Gamma^{\mu}X\partial_{\mu} \Psi_{+})] \nonumber  .
\end{eqnarray}
Some remarks are in order. The first line refers to full spinors, while
the suffices ``$//$" and ``$\bot$" are used to denote bilinear terms
constructed with Weyl spinors of,
respectively, same and opposite chiralities. Please notice that,
due to the non-associativity of the octonions, we need to specify the correct
order in which the operations are taken (this is not necessary in the quaternionic 
case). The symbol ``$X$" denotes the possibility of introducing, depending on
the given space-time, external, extra-type of Gamma matrices, which could be
either of time-like, or of space-like nature. More specifically, in the tables below,
$X$ will be denoted as $T$, $S$, $J$ or $F$ whether it will be associated with
external time-like Gamma matrices ($T$), space-like ($S$) ones, the product of two of them
($J$) or, finally, of three of them ($F$). The presence of an extra-number
specifies how many inequivalent choices for the introduction of such matrices can be given.\par
It should also be noticed that, in the octonionic case, the symbol ``$tr$" introduced on the r.h.s. denotes the projection over the octonionic identity (while in the quaternionic case it
coincides with the usual definition of the trace).\par
In full analogy with the (pseudo)-kinetic terms, (pseudo)-massive terms in
a lagrangian action can be introduced through
\begin{eqnarray}
 M_{X} &=& tr(\Psi^{\dagger}_{} A
X \Psi_{}), \nonumber \\
 M_{//X} &=& tr(\Psi^{\dagger}_{+} A
X \Psi_{+}), \nonumber \\ M_{\bot X} &=&
tr(\Psi^{\dagger}_{+} A X\Psi_{-} + \Psi^{\dagger}_{-}
A X\Psi_{+}),\nonumber 
\label{bilinmass}
\end{eqnarray}
Due to the anticommuting character of the spinors and of their basic components, the 
non-vanishing (pseudo)kinetic and (pseudo)massive terms are only allowed in
given spacetimes. In the next two subsection we report the full classification
of the allowed Dirac equations in, respectively, the quaternionic and octonionic cases.

\subsection{The quaternionic Dirac equations}

We present here the tables of the allowed free (pseudo)-kinetic and
(pseudo)-massive terms for quaternionic spinors. The columns are labeled by 
$t~mod~4$ and the rows by
$t-s~mod~8$ ($t,s$ denoting the number of time-like and space-like directions
of the given space-time), while the symbols used in the entries have been explained above.\par For full spinors ($NW$) case we have
\begin{eqnarray}&&
{\begin{tabular}{||c||c|c|c|c|c|} \hline \hline
   & 0 & 1 & 2 & 3 \\ \hline
    \hline
   0&$K_{J_j}$, $K_F$, $M_{S_j}$, $M_{J_j}$  &$K$, $K_F$, $M_{J_j}$&
$K$, $K_{S_j}$, $M$, $M_F$ &$K_{S_j}$, $K_{J_j}$, $M$, $M_{S_j}$,
$M_F$ \\ \hline
  5  &  &  $K$ & $K$, $M$ &  $M$  \\ \hline
  6 & $M_{S}$  & $K$  & $K$, $K_{S}$, $M$ &  $K_{S}$, $M$, $M_{S}$  \\  \hline
  7 &$ K_{J}$, $ M_{S_i}$,$ M_{J}$ & $K$, $ M_{J}$ &
  $K$, $K_{S_i}$, $M$ & $K_{S_i}$ ,$ K_{J}$, $M$, $M_{S_i}$ \\ \hline

 \end{tabular}}\nonumber\end{eqnarray}
\begin{eqnarray}
&&
\end{eqnarray}
while for Weyl spinors ($W$ case) we get
\begin{eqnarray}&&
{\begin{tabular}{||c||c|c|c|c|c|} \hline \hline
   & 0 & 1 & 2 & 3 \\ \hline
    \hline
  1&$
\begin{array}{l} K_{// T_j}, K_{\bot J_j},\\ M_{//J_j}
\end{array}$& $\begin{array}{l} K_{//},
K_{\bot T_j},\\ M_{// F}, M_{//J_j}, M_{\bot J_j}\end{array}$&
$\begin{array}{l}K_{// F}, K_\bot,\\ M_{//}, M_{\bot F}, M_{\bot
T_j}\end{array}$& $\begin{array}{l} K_{\bot F}, K_{// J_j},\\
M_\bot \end{array}$
\\ \hline
  2 & $\begin{array}{l} K_{// T_i}, K_{\bot J},\\ M_{// J}\end{array} $ &
$\begin{array}{l} K_{//}, K_{\bot T_i},\\
  M_{// T_i}, M_{\bot J}\end{array}$ &
$\begin{array}{l} K_{\bot},\\ M_{//},M_{\bot T_i}\end{array}$  &
$\begin{array}{l} K_{// J},\\ M_{\bot}\end{array}$   \\ \hline
  3  & $\begin{array}{l}K_{//T}\\ \end{array}$ & $\begin{array}{l}K_{//}, K_{\bot T},\\
 M_{//T}\end{array}$   & $\begin{array}{l} K_\bot , \\M_{//}, M_{\bot T}\end{array}$
  & $\begin{array}{l} \\ M_{\bot}\end{array}$    \\  \hline
  4 &  & $\begin{array}{l} K_{//}\\ \end{array}$  & $\begin{array}{l} K_\bot , \\
M_{//}\end{array}$  &  $\begin{array}{l}\\M_{\bot}\end{array}$
\\  \hline
 \end{tabular}}\nonumber\end{eqnarray}
\begin{eqnarray}
&&
\end{eqnarray}
Please notice that in the two tables above the suffix ``$j$"
denotes the existence of three inequivalent choices for the
corresponding matrices (e.g., the three distinct space-like
matrices $S_j$), while the suffix ``$i$" denotes the existence of
two inequivalent choices. 

\subsection{The octonionic Dirac equations.}

In full analogy with the previous case, we are able to produce the tables
corresponding to the allowed (pseudo)-kinetic and (pseudo)-massive terms
in octonionic Dirac equations. We get, in the ($NW$) case
\begin{eqnarray}&&
{\begin{tabular}{||c||c|c|c|c|c|} \hline \hline
   & 0 & 1 & 2 & 3 \\ \hline
    \hline
  1  &  &  $K$ & $K$, $M$ &  $M$  \\ \hline
  2 & $M_{S}$  & $K$  & $K$, $K_{S}$, $M$ &  $K_{S}$, $M$, $M_{S}$  \\  \hline
  3 & $ K_{J}$,$ M_{S_i}$,$ M_{J}$ & $K$, $ M_{J}$ &
  $K$, $K_{S_i}$, $M$ & $K_{S_i}$, $ K_{J}$, $M$, $M_{S_i}$ \\ \hline
4 & $K_{J_j}$, $K_F$, $M_{S_j}$, $M_{J_j}$& $K$, $K_F$, $M_{J_j}$,
$M_F$& $K$, $K_{S_j}$, $M$, $M_F$& $K_{S_j}$, $K_{J_j}$, $M$,
$M_{S_j}$\\ \hline
 \end{tabular}}\nonumber\end{eqnarray}
\begin{eqnarray}
&&
\end{eqnarray}
while in the $W$ (Weyl) case we obtain
\begin{eqnarray}&&
{\begin{tabular}{||c||c|c|c|c|c|} \hline \hline
   & 0 & 1 & 2 & 3 \\ \hline
    \hline
  0 &  & $\begin{array}{l} K_{//}\\ \end{array}$  & $\begin{array}{l}
K_\bot ,\\ M_{//}\end{array}$ &$\begin{array}{l} \\
M_{\bot}\end{array}$   \\  \hline 5 & $\begin{array}{l} K_{//T_j},
K_{\bot J_{j}},\\ M_{// J_j}, M_{\bot F} \end{array}$ &
$\begin{array}{l} K_{//}, K_{\bot T_j}, \\ M_{// T_j}, M_{\bot
J_j}\end{array}$& $\begin{array}{l} K_{\bot} , K_{//F}, \\ M_{//},
M_{\bot T_j}\end{array}$& $\begin{array}{l} K_{// J_j}, K_{\bot
F}, \\ M_{\bot}, M_{// F} \end{array}$\\ \hline
  6 & $\begin{array}{l}K_{// T_i}, K_{\bot J},\\  M_{// J},\end{array}$  & $\begin{array}{l}
  K_{//}, K_{\bot T_i}, \\
  M_{// T_i}, M_{\bot J}\end{array}$ & $\begin{array}{l} K_\bot , \\M_{//},
  M_{\bot T_i}\end{array}$  & $\begin{array}{l} K_{// J},\\ M_{\bot}\end{array}$   \\
  \hline
  7  & $\begin{array}{l}K_{//T}\\
\end{array}$ & $\begin{array}{l}K_{//},K_{\bot T},\\ M_{//T}\end{array}$  &
$\begin{array}{l} K_\bot ,\\M_{//}, M_{\bot T}\end{array}$
 & $\begin{array}{l} \\M_{\bot}\end{array}$    \\  \hline
 \end{tabular}}\nonumber\end{eqnarray}
The introduced symbols have the same meaning as before.

\section{Quaternionic and octonionic supersymmetries}

It comes to no surprise that the most potentially interesting physical applications
of the formalism and results previously introduced concern supersymmetry. After all,
supersymmetry is nowadays (in the superstring/M-theory scenario) a necessary
ingredient for our present understanding of fundamental interactions.
\par
If octonions should play any role at all in physics, this is quite likely being in relation
with the unification of all interactions realized by supersymmetry in higher-dimensional
spacetimes. The mere fact that an octonionic formulation of the $M$-algebra is available
\cite{LT1} gives us some hopes that this program could one day be carried out.\par
In this section we will briefly review how generalized supersymmetries in higher-dimensional
space-times (which are regarded as the scenarios for the unification of interactions and must be supplemented by
some dimensional-reduction mechanism such as the Kaluza-Klein compactifications) can be
constructed in terms of quaternionic and octonionic spinors (depending on the spacetime,
either within or without a Weyl
projection). The scheme of this section is therefore the following, at first the necessary
ingredients to define (division-algebra valued) generalized supersymmetries are introduced.
Next, the two cases of quaternionic and octonionic supersymmetries are more closely
analyzed. We especially focus on  results concerning the $D=11$-dimensional spacetimes, 
mostly because this is the space-time dimensionality where the supposed $M$-theory should live.
\par
In terms of $n$-component real spinors $Q_a$, the most general real supersymmetry algebra is represented by
\begin{eqnarray}\label{Mgen}
    \left\{ Q_a, Q_b \right\} & = & {\cal Z}_{ab},
\end{eqnarray}
where the matrix ${\cal Z} $ appearing in the r.h.s. is the most general $n\times n$
symmetric matrix with total number of $\frac{n(n+1)}{2}$ components. For any given space-time we
can easily compute its associated decomposition
of ${\cal Z}$ in terms of the antisymmetrized products of $k$-Gamma matrices, namely
\begin{eqnarray}
{\cal Z}_{ab} &=& \sum_k(A\Gamma_{[\mu_1\ldots\mu_k]})_{ab}Z^{[\mu_1\ldots \mu_k]},
\end{eqnarray}
where the values $k$ entering the sum in the r.h.s. are restricted by the symmetry requirement for the 
$a\leftrightarrow b$ exchange
and are specific for the given spacetime. The coefficients $Z^{[\mu_1\ldots \mu_k]}$ are the rank-$k$ abelian tensorial central charges.\par
In the Minkowskian $(10,1)$ space-time, supporting $32$-component real spinors, the
bosonic r.h.s. is split into the $11+55+462=528$ bosonic components sectors $M_1+M_2+M_5$, where the $k$ in $M_k$, for
$k=1,2,5$, specifies the level of the rank-$k$ antisymmetric tensors.
  \par
When the the fundamental spinors entering the supersymmetry algebra belong to a division algebra
other than the real one (this is evidently true in the quaternionic and octonionic cases),
an extra possibility is available. The most general supersymmetry algebra can be
expressed in terms of 
 anticommutators among the fundamental
spinors $Q_a$ and their conjugate ${Q^\ast}_{\dot a}$, where the conjugation refers to the principal
conjugation in the given division algebra. One should remember that the principal conjugation,
restricted to real spinors, acts as the identity, see (\ref{conjug}).
In the quaternionic and octonionic (as well as complex) cases we have
\begin{eqnarray}\label{Mhol}
    \left\{ Q_a, Q_b \right\} =  {\cal Z}_{ab}\quad &,& \quad \left\{ {Q^\ast}_{\dot a}, {Q^\ast}_{\dot b} \right\} =  {{\cal Z}^\ast}_{\dot{a}\dot{b}},
\end{eqnarray}
together with
\begin{eqnarray}\label{Mher}
\left\{ Q_a, {Q^\ast}_{\dot b} \right\} &=&  {\cal W}_{{a}\dot{b}},
\end{eqnarray}
where the matrix ${\cal Z}_{ab}$ (${{\cal Z}^\ast}_{\dot{a}\dot{b}}$ is its conjugate and does not contain new degrees of freedom) is symmetric,
while ${\cal W}_{{a}\dot{b}}$ is hermitian.\par
Two big classes of subalgebras, respecting the Lorentz-covariance,
can be obtained from (\ref{Mhol}) and (\ref{Mher}) by imposing division-algebra
constraints, obtained by setting identically equal to zero either ${\cal Z}$ or ${\cal W}$, namely ${\cal Z}_{ab}\equiv   {{\cal Z}^\ast}_{\dot{a}\dot{b}}\equiv 0$, so that the only bosonic degrees
of freedom enter the hermitian matrix ${{\cal W}}_{{a}\dot{b}}$ or, conversely,
${{\cal W}}_{{a}\dot{b}}\equiv 0$, so that the only bosonic degrees of freedom enter 
${\cal Z}_{ab}$ and its conjugate matrix ${{\cal Z}^\ast}_{\dot{a}\dot{b}}$.
The first type of constraint will be referred as the one giving rise to the ``hermitian" generalized supersymmetries, while the
generalized supersymmetries satisfying the second constraint will be referred to
as ``holomorphic" generalized supersymmetries.\par
Several other constraints can be imposed, for instance one can consistently set,
for complex spinors, the matrix $Z$ entering (\ref{Mhol}) to be real. However, for our 
purposes, it is enough to concentrate on hermitian and holomorphic supersymmetries.

\subsection{Quaternionic supersymmetries}

Both the hermitian and holomorphic quaternionic supersymmetries can be classified 
with the help of tables specifying the number and type of bosonic elements (abelian tensorial central charges of rank $k$) entering the r.h.s. . It is worth noticing that the results do not depend on the signature of the associated space-time, but only on its dimensionality
$D$, provided of course that the associated spacetime is actually carrying quaternionic spinors.\par
For what concerns the quaternionic hermitian supersymmetry we get
{ {{\begin{eqnarray}&\label{hh1}
\begin{tabular}{|c|c|c|}\hline
spacetime&bosonic sectors&bosonic components\\ \hline
$D=3$&${M}_0$& $1$\\ \hline

$D=4$&${M}_0$& $1$\\ \hline

$D=5$&${M}_0+{M}_1$&$1+5=6$\\ \hline

$D=6$&${M}_1$&$6$\\ \hline

$D=7$&${M}_1+{M}_2$& $7+21=28$\\ \hline

$D=8$&${M}_2$&$28$\\ \hline

$D=9$&${M}_2+{M_3}$&$36+84=120$\\ \hline

$D=10$&${M}_3$&$120$\\ \hline

$D=11$&${M}_0+{ M}_3+{ M}_4$&$1+165+330=496$\\ \hline

$D=12$&${M}_0+{M}_4$&$1+495=496$\\ \hline

$D=13$&${M}_0+{M}_1+{M}_4+{M}_5$&$1+13+715+1287=2016$\\ \hline

\end{tabular}&\end{eqnarray}}} }  
The last column denotes the number of bosonic components enetring the rank-$k$ decomposition.
It is worth noticing that the hermitian quaternionic supersymmetry
saturates the bosonic sector.\par
This property is not hold by the holomorphic quaternionic supersymmetry.
The reason can be traced to the fact that
if we try implementing transposition on imaginary quaternions we are in conflict with their
product since, e.g., $(e_1\cdot e_2)^T={e_2}^T\cdot {e_1}^T = - e_3 \neq {e_3}^T$. Indeed the only consistent operation respecting the composition law 
would correspond to setting ${e_i}^T=-{e_i}$, but this in fact coincides with the principal conjugation employed
in the construction of quaternionic hermitian matrices and quaternionic hermitian supersymmetries.
The holomorphic analog of the previous table is given by
{ {{\begin{eqnarray}&\label{hh11}
\begin{tabular}{|c|c|c|}\hline
spacetime&bosonic sectors&bosonic components\\ \hline
$D=3$&${ M}_0+{ M}_1$& $1+3=4$\\ \hline

$D=4$&${{ M}}_1$& $4$\\ \hline

$D=5$&${{ M}}_1$&$5$\\ \hline

$D=6$&$-$&$-$\\ \hline

$D=7$&$-$&$-$\\ \hline

$D=8$&$-$& $-$\\ \hline

$D=9$&${M}_0$&$1$\\ \hline

$D=10$&${M}_0+{{M}}_1$&$1+10=11$\\ \hline

$D=11$&${M}_0+{M}_1$&$1+11=12$\\ \hline

$D=12$&${M}_1$&$12$\\ \hline

$D=13$&${M}_1$&$13$\\ \hline

\end{tabular}&\end{eqnarray}}} }  
The above results can be interpreted as follows: quaternionic holomorphic supersymmetry
cannot admit bosonic tensorial central charges of rank $k\geq 2$. At most a single bosonic
central charge ($M_0$), depending on the dimensionality of the space-time, can exist. In some
dimensions, no consistent quaternionic holomorphic supersymmetry can be defined. \par
From physical point of view, so far, the most interesting application of supersymmetries of quaternionic spacetimes does not directly concern the quaternionic supersymmetry, but 
the complex holomorphic supersymmetry which can be realized with the quaternionic spinors entering the $11$-dimensional quaternionic spacetime $(0,11)$. The bosonic components correspond to the $528$ bosonic components of the real $M$ algebra and the $11D$ complex Euclidean holomorphic supersymmetry can be regarded as the Euclideanized version of the
$M$ algebra, see \cite{LT3}.

\subsection{Octonionic supersymmetries}

Let us discuss now the peculiar features of the octonionic supersymmetries which are consequences of the non-associativity of octonions.
The octonionic supersymmetries exist for the spacetimes entering the (\ref{octonspaces}) table. 
It is worth mentioning that here we limit ourselves to consider only ``hermitian" octonionic
supersymmetries. \par
In a $D$-dimensional spacetime described in terms of the octonions, $D-7$ Clifford Gamma matrices are purely real, while the remaining $7$ of them are given by the imaginary
octonions $e_i$, $i=1,2,\ldots ,7$, multiplying a common real matrix.
In describing the antisymmetric product of $k>2$ octonionic $\Gamma$-matrices
a correct prescription must be specified to take into account the
non-associativity of the octonions. As a matter of fact, the correct prescription
can be induced by assuming a given prescription for the antisymmetrized product of $k>2$
imaginary octonions $e_i$. The correct prescription can be uniquely specified by assuming
the validity of the Hodge duality and an irreducibility
requirement, namely that the rank-$k$ antisymmetric product of $k$ imaginary octonions are
either proportional to the octonionic identity or to the imaginary octonions. In full generality, this prescription corresponds at taking the following antisymmetrized product
of $k$ octonionic Gamma matrices  
\begin{eqnarray}
\relax [\Gamma_{1}\cdot \Gamma_{2}\cdot \dots \cdot \Gamma_k]
&\equiv& \frac{1}{k!}\sum_{perm.} (-1)^{\epsilon_{i_1\dots i_k}}
(\Gamma_{i_1}\cdot \Gamma_{i_2}\dots \cdot \Gamma_{i_k}),
\label{antiprod}
\end{eqnarray}
where $(\Gamma_1\cdot \Gamma_2\dots \cdot \Gamma_k)$ denotes the
symmetric product
\begin{eqnarray}
(\Gamma_1\cdot \Gamma_2 \cdot\dots  \cdot \Gamma_k) &\equiv&
\frac{1}{2}(. ((\Gamma_1 \Gamma_2)\Gamma_3\dots )\Gamma_k)
+\frac{1}{2} (\Gamma_1(\Gamma_2(\dots \Gamma_k)).).
\end{eqnarray}
The usefulness of this prescription is due to the fact that the
product \begin{eqnarray}& A\relax [\Gamma_{1}\cdot \Gamma_{2}\cdot
\dots \cdot \Gamma_k],& \label{agammas}
\end{eqnarray}
(where $A$ is the matrix, product of the time-like Gamma matrices,
already introduced at the beginning of this section) has a definite (anti)-hermiticity
property. The different (\ref{agammas}) tensors, for different
choices of the Gamma's, are all hermitian or antihermitian,
depending only on the value of $k$ and not of the $\Gamma$'s
themselves.
In odd-dimensions $D$ we get the table, whose columns are
labeled by the antisymmetric tensors rank $k$, 
specifying the number of independent bosonic components in each rank-$k$
antisymmetric product (\ref{antiprod}). 

 {{\begin{eqnarray}&
\begin{tabular}{|c|c|c|c|c|c|c|c|c|c|c|c|c|c|c|}\hline
&$0$&$1$&$2$&$3$&$4$&$5$&$6$&$7$&$8$&$9$&$10$&$11$&$12$&$13$\\
\hline $D=7$&${\underline {1}}$&$7$&$7$&${\underline
{1}}$&${\underline{ 1}}$&$7$&$7$&${\underline{ 1}}$&&&&&&\\ \hline

$D=9$&${\underline {1}}$&${\underline {9}}$&$22$&$22$&$
{\underline {10}}$&${\underline {10}}$&$22$&$22$&${\underline
{9}}$&${\underline {1}}$&&&&\\ \hline

$D=11$&$1$&${\underline {11}}$&${\underline
{41}}$&$75$&$76$&${\underline {52}}$&${\underline
{52}}$&$76$&$75$&${\underline {41}}$&${\underline {11}}$&$1$&&\\
\hline

$D=13$&$1$&$13$&${\underline {64}}$&${\underline{
168}}$&$267$&$279$&${\underline{ 232}}$&${\underline{
232}}$&$279$&$267$&${\underline{ 168}}$&${\underline{
64}}$&$13$&1\\ \hline

\end{tabular}&\nonumber\end{eqnarray}}}
\begin{eqnarray}
&&
\end{eqnarray}
An analogous table can be produced in even-dimensional spacetimes as well.\par
In the above table the $k$ sectors corresponding to hermitian matrices (and therefore
entering the r.h.s. of a generalized supersymmetry) are underlined.\par
The table above shows the existence of identities relating
higher-rank antisymmetric octonionic tensors. Let us discuss the 
$D=11$ example. The $52$ independent components of an
octonionic hermitian $(4\times 4)$ matrix can be expressed either
as a rank-$5$ antisymmetric tensors (simbolically denoted as
``$M5$"), or as the combination of the $11$ rank-$1$ ($M1$) and
the $41$ rank-$2$ ($M2$) tensors. The relation between $M1+M2$ and
$M5$ can be made explicit as follows. The $11$ vectorial indices
$\mu$ are split into $4$ real indices, labeled by $a,b,c,\ldots$
and $7$ octonionic indices labeled by $i,j,k,\ldots$. We get, on
one side, {{\begin{eqnarray}&
\begin{tabular}{cc}

$4$& $M1_a$\\

$7$&$M1_i$\\

$6$&$M2_{[ab]}$\\

$4\times 7= 28$&$M2_{[ai]}$\\

$7$& $ M2_{[ij]}\equiv M2_{i}$

\end{tabular}&\nonumber\end{eqnarray}}}

while, on the other side, {{\begin{eqnarray}&
\begin{tabular}{cc}

$7$& $M5_{[abcdi]} \equiv M5_i$\\

$4\times 7=28$&$M5_{[abcij]}\equiv M5_{[ai]}$\\

$6$&$M5_{[abijk]}\equiv M5_{[ab]}$\\

$4$&$M5_{[aijkl]}\equiv {M5}_a$\\

$7$& $ M5_{[ijklm]}\equiv {\widetilde M5}_{i}$

\end{tabular}&\nonumber\end{eqnarray}}}
which shows the equivalence of the two sectors, as far as the
tensorial properties are concerned. Please notice that the correct
total number of $52$ independent components is recovered
\begin{eqnarray}
52 &=& 2\times 7 +28+6+4.
\end{eqnarray}
The octonionic equivalence of different antisymmetric tensors can
be symbolically expressed, in odd space-time dimensions, through
{{\begin{eqnarray}&
\begin{tabular}{|c|c|}\hline

$D=7$& $M0\equiv M3$\\ \hline

$D=9$&$M0+M1\equiv M4$\\ \hline

$D=11$&$M1+M2\equiv M5$\\ \hline

$D=13$&$M2+M3\equiv M6$\\ \hline

$D=15$&$M3+M4\equiv M0+M7$\\ \hline

\end{tabular}&\label{tablem}\end{eqnarray}}}

\subsection{The octonionic $M$-algebra}

We are now in the position to introduce the octonionic $M$ algebra \cite{LT1}.
\par
It corresponds to replace the real supersymmetry algebra in the $(10,1)$
spacetime, given by
\begin{equation}
\{Q_a,Q_b\}= (A\Gamma_\mu )_{ab}P^\mu +(A\Gamma_{[\mu\nu]})_{ab}
Z^{[\mu\nu]} +
(A\Gamma_{[\mu_1\dots\mu_5]})_{ab}Z^{[\mu_1\dots\mu_5]}
\label{Malg}
\end{equation}
with its two octonionic-valued variants, given by $4$-component octonionic spinors
$Q_a$ (together with their conjugate spinors
${Q^\ast}_{b} $) and the
$52$ octonionic-valued $4\times 4$ hermitian matrices
which can be expressed, either as the $11+41$ bosonic generators entering
\begin{equation}\label{eq1}
 \left\{ Q_a, {Q^\ast}_{b} \right\} = P^\mu (A\Gamma^{}_\mu)_{ab} +
   Z^{\mu\nu}_{\bf{O}} (A\Gamma^{}_{\mu\nu})_{ab}
   ,
\end{equation}
or as the $52$ bosonic generators entering
\begin{equation}\label{eq2}
 \left\{ Q_a, {Q^\ast}_{b} \right\} =
    Z^{[\mu_1\ldots \mu_5]}_{\bf{O}}
    (A\Gamma^{}_{\mu_1 \ldots
    \mu_5})_{ab}\, .
\end{equation}
Associated to the above octonionic $M$ algebra, its superconformal extension, 
given by the $Osp(1,8|{\bf O})$ superalgebra, can be constructed \cite{LT2}.

\section{Conclusions}

In this paper we have quickly reviewed the fundamental issues concerning the 
employment of quaternionic and octonionic spinors. In particular we have
described a general construction which allows us to specify the spacetimes supporting
quaternionic and octonionic spinors. The specific problems raised by the non-associativity
of the octonions have been discussed and clarified. It was shown, in particular, that
octonionic spinors naturally encode the geometry of the Euclidean seven-sphere $S^7$.\par
With our tools we were able to construct and classify all free Dirac-type equations
involving quaternionic and octonionic spinors. The concepts of Weyl spinors, the presence
of (pseudo)-kinetic and (pseudo)-massive terms in association with different space-times
have been fully investigated and the complete list of results has been reported.\par
In the last part of this talk we took a further step. In view of studying the possible
physical consequences of the formalism here introduced we applied the above investigation
to the construction and the classification of the generalized supersymmetries supported by
quaternionic and octonionic spinors. Some recent results on this subject have been reported,
like the notion of division-algebra constrained (in the quaternionic case) hermitian and
holomorphic suprsymmetries. \par
By far the most intriguing possible application of the ideas related with the octonionic spinors concern the $M$-theory investigations, as suggested in \cite{LT1} and \cite{LT2}.
Indeed, it is quite remarkable that an octonionic structure can be introduced, instead of
the standard real structure, in defining the (octonionic version of) the $M$-algebra.   
Peculiar identities, relating different rank-$k$ antisymmetric tensors of the bosonic
sectors, are an absolute novel feature of the octonionic formulation, finding no counterparts in the standard formulation. It is worth noticing that, in a somewhat different context,
octonions have been suggested \cite{Boy} to be linked to a possible exceptional formulation
\cite{Ram} for a single unifying theory of all interactions.\par
The investigations concerning the dynamics of octonionic spinors are a necessary preliminary
step to unveil this challenging and fascinating present area of research. 

%x\cite{Brown2000,BrownAustin:2000}
%\cite{Knuth:WEB}
%%%%%%%%%%%%%%%%%%%%%%%%%%%%%%%%%%%%%%%%%%%%%%%%
%% BACKMATTER
%%%%%%%%%%%%%%%%%%%%%%%%%%%%%%%%%%%%%%%%%%%%%%%%

\begin{theacknowledgments}
The present paper reports results based on a series of works 
done in collaboration with
J. Lukierski and with  H.L. Carrion and M. Rojas, who I am pleased to acknoweldge.
\end{theacknowledgments}

%% If the bibliography is produced without BibTeX comment out the


%\bibliography{toppan}

\begin{thebibliography}{9}
\bibitem{Por} Porteous, I.R., \emph{Clifford Algebras and the Classical Groups}, 
Cambridge Un. Press, 1995.
\bibitem{Oku} Okubo, S., \emph{J. Math. Phys.} {\bf 32}, 1657 (1991); {\em ibid.}
{\bf 32}, 1669 (1991).
\bibitem{CRT} Carrion, H.L., Rojas, M. and Toppan, F.,
JHEP04 (2003) 040.
\bibitem{KT} Kugo, T. and Townsend, P., \emph{Nucl. Phys.} {\bf B 221}, 357
(1983).
\bibitem{Bae} Baez, J., \emph{The Octonions}, math.RA/0105155.
\bibitem{BS} Barton, C.A. and Sudbery, T., math.RA/0203010.
\bibitem{GG} G\"{u}naydin, M. and G\"{u}rsey, F., \emph{Lett. Nuovo Cim.}
{\bf 6} (1973), 401.
\bibitem{Boy} Boya, L., \emph{Octonions and M-theory}, hep-th/0301037.
\bibitem{FM} Fairlie, D.B. and Manogue, A.C., \emph{Phys. Rev.} {\bf D 34},
1832 (1986).
\bibitem{LT1} Lukierski, J. and Toppan, F., \emph{Phys. Lett.} {\bf B 539},
266 (2002).
\bibitem{LT2} Lukierski, J. and Toppan, F., \emph{Phys. Lett.} {\bf B 567},
125 (2003).
\bibitem{Ram} Ramond, P., \emph{Algebraic Dreams}, hep-th/0112261.
\bibitem{Top} Toppan, F., JHEP09 (2004) 016.
\bibitem{LT3} Lukierski, J. and Toppan, F.,\emph{Phys. Lett.} {\bf B 584},
315 (2004).
\bibitem{LM} Lukierski, J. and Minnaert, P., \emph{Phys. Lett.} {\bf B 129}
(1983), 392.
\end{thebibliography}

%  \typeout{******************************************}
%  \typeout{** Please run "bibtex \jobname" to optain}
%  \typeout{** the bibliography and then re-run LaTeX}
%  \typeout{** twice to fix the references!}
%  \typeout{******************************************}

\end{document}